\documentclass[prb,showpacs,twocolumn,amsmath,amssymb]{revtex4}
\usepackage{epsfig}
\usepackage{graphicx}
\usepackage{dcolumn}
\usepackage{bm}
\usepackage{appendix}

\newcommand{\vk}{{\bf k}}

\newcommand{\vq}{{\bf q}}

\newcommand{\vp}{{\bf p}}

\renewcommand{\vr}{{\bf r}}

\begin{document}
\title{Quantum waveguide theory of Andreev spectroscopy in multiband superconductors: the
case of Fe-pnictides}
\author{M. A. N. Ara\'ujo$^{1,2}$, P. D. Sacramento$^{1}$}
\affiliation{$^1$ CFIF, Instituto Superior 
T\'ecnico, UTL, Av. Rovisco Pais, 1049-001 Lisboa, Portugal}
\affiliation{$^2$ Departamento de F\'{\i}sica,  Universidade de \'Evora, P-7000-671, \'Evora, Portugal}

\begin{abstract}
The problem of Andreev reflection between a normal metal and a multiband superconductor 
is addressed. The appropriate matching conditions for the wave function at the interface are established 
on the basis of an  extension of quantum waveguide theory to these systems. 
Interference effects between  different bands of the superconductor  manifest themselves in the conductance and  
the case of FeAs superconductors is specifically considered, in the framework of a recently proposed 
effective two-band model, in the  sign-reversed  s-wave pairing scenario.   
Resonant transmission through surface  Andreev bound states is found as well as
destructive interference effects  that produce zeros in the conductance at normal incidence. 
Both these effects occur at nonzero bias voltage. 
\end{abstract}
\pacs{73.20.-r,74.20.Rp, 74.50.+r,74.70.Dd}

\maketitle

\section{Introduction}

Electronic scattering at the interface between a normal metal (N) 
and a superconductor has been  used as a probe to investigate the 
electronic properties of superconductors\cite{tinkham,greene} and, 
more recently, FeAs superconductors(FAS), leading, in the latter case,  
to different conclusions regarding the  pairing symmetry\cite{tesanovic,chineses}.
As compared to conventional and high-T$_c$ materials,
the recently discovered FeAs based superconductors have a more complex band structure, 
with a Fermi surface (FS) consisting of four sheets, two of them hole-like, and the other 
two electron-like\cite{singh,xu,mazin,haule}. S-wave, d-wave and p-wave pairing scenarios 
have been proposed to describe the superconducting state\cite{chubukovs,qi,lee,yao}.
One of the suggested pairing scenarios
is the so-called sign-reversed s-wave state (s$^\pm$-state), where the gap function has opposite
signs in the hole-like and the electron-like sheets of the FS. 
Since this is a novel possibility, it deserves some theoretical development.
A recent experiment seems to confirm this pairing scenario in a 122 compound\cite{Christianson}.

Blonder {\it et al}\cite{blonder} devised a theory for Andreev scattering in isotropic s-wave superconductors
which has been later generalized to unconventional (anisotropic) superconductors \cite{tanaka}.
These theories  apply to one band  superconductors.
In the case of  multiband superconductors (MBS), such as FAS and heavy-fermion compounds, the 
bands are usually treated as separate conduction channels with (classically) additive conductances\cite{sudbo},
like paralel resistors, thereby neglecting the quantum mechanical nature of  the scattering problem 
at the interface, where interference effects between the transmitted waves in  different bands of the MBS are expected.
Such interference effects will lead to new features in the conductance.

We are thus posed the problem of finding the wave function for the scattering  state of an incident particle
from a one-band metal which  is transmitted through two or more bands inside the superconductor.
The splitting of the incident electron's probability amplitude among several conduction channels
is the same quantum mechanical problem as in a quantum waveguide.
Thus, in order to derive the appropriate matching conditions for the wave function at the interface, we need to
make an extension of quantum waveguide theory.

Applying to the case of FAS, we obtain the differential conductance curves {\it vs} 
bias voltage and explicitly show  the emergence of Andreev bound states (ABS)
in the s$^\pm$-state scenario, as a manifestation of interference effects between the bands,
unlike the usual ABS in one-band superconductors. An unusual feature of the ABS is that they  occur at 
a finite energy above the Fermi level and disperse with the electron's transverse momentum. On the other hand,
interference effects may also suppress the conductance at certain energies.

\section{Quantum waveguide theory}

\begin{figure}[htb]
\centerline{\includegraphics[width=6.5cm]{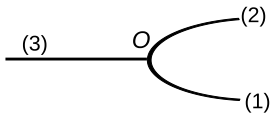}}
\caption{\label{waveguide} 
Three branches of a waveguide with a node at $O$.}
\end{figure}

The splitting of the incident electron's probability amplitude among several conduction channels
is the same quantum mechanical problem as in a quantum waveguide.
In a  quantum waveguide,  three one-dimensional
conductors intercept at one point  (see Figure \ref{waveguide})\cite{xia}. 
The wavefunction for a particle must be continuous and  single-valued at
the circuit node $O$, implying that
\begin{equation}
\psi(x_1\rightarrow O) =\psi(x_2\rightarrow O)= \psi(x_3\rightarrow O)\,,
\label{um}
\end{equation}
where $x_1, x_2, x_3$ are coordinates along branches 1, 2 and 3 respectively.
The (probability and charge) current conservation at the node is guaranteed by the "quantum Kirchhoff" law\cite{xia}
\begin{equation}
\sum_{j=1}^3 \frac{1}{m_j}\frac{\partial\psi(x_j\rightarrow O)}{\partial x_j} = 0\,,
\label{dois}
\end{equation}
where the coordinates $x_j$ ($j=1,2,3$) must be all of them directed to (or away from) the node $O$
and $m_j$ denotes the particle's effective mass in branch $j$. 

\begin{figure}[htb]
\centerline{\includegraphics[width=8.0cm]{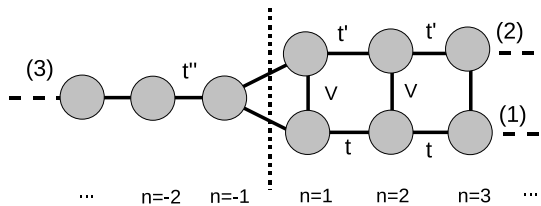}}
\caption{\label{tightexample} 
Tight-binding waveguide  with three branches. In branches 1 and 2 there is electron hopping along ($t$,$t'$) and perpendicular ($V$) to the chains.
An integer $n$ labels unit cells along the chain. 
}
\end{figure}

A simple one-dimensional version of the  N/MBS interface is a tight-binding  chain which  has a bifurcation at some point, 
as shown in Figure \ref{tightexample}. We further assume 
the sites in branch 1  to be  coupled to branch 2 through a hybridization operator, $\hat V$. 
An integer $n$ labels the two-atom unit cell along the chain. 

Let $|n,j\rangle$ denote the site in cell $n$ of chain $j$. Then, the incoming particle in branch 3 with wavevector $p$ 
is described by the wavefunction $\psi_{inc}(n) = e^{ipn} + b\ e^{-ipn}$, where $b$ denotes the reflection amplitude.

If chains 1 and 2 were decoupled,  a Bloch state in chain $j$ would have momentum $k$ and energy $\epsilon_j(k)$. 
But now suppose that an operator $\hat V$ hybridizes Bloch states in the two chains. 
The Hamiltonian matrix for the coupled chains 1+2,  $\hat H_{1+2}$, has an off-diagonal element, $V(k)$,
and its eigenstates  follow from  the eigenproblem: 
\begin{equation}
\left(\begin{array}{cc} \epsilon_1(k)& V(k)\\ V(k)&\epsilon_2(k)\end{array}\right) \left( \begin{array}{c}\alpha\\ \beta\end{array}\right) = 
{\cal E}\left(\begin{array}{c}\alpha\\ \beta\end{array} \right)\,,
\label{eigen}
\end{equation}
which yields two bands,  ${\cal E}_\pm(k)$, so that a Bloch state in the coupled chains 1+2 has the form
$$
\phi(n) = \left(\begin{array}{c}\alpha_k \\ \beta_k\end{array}\right)e^{ikn}\,.
$$
The eigenvector components, $\alpha$ and $\beta$, 
denote the wavefunction projections on branches 1 and 2, respectively.
The wavefunction for the transmitted particle in chains 1+2 reads, for $n>0$,
\begin{eqnarray}
\psi_t(n) &=& C \left(\begin{array}{c}\alpha_k\\ \beta_k\end{array}\right) e^{ikn}
+ D\left(\begin{array}{c}\alpha_{k'}\\ \beta_{k'}\end{array}\right) e^{ik'n} \,,
\label{transmit}
\end{eqnarray}
where the momenta satisfy the energy conservation condition
 ${\cal E}_-(k)={\cal E}_+(k')=\epsilon(p)$.
We now  join the wavefunction in branch 3 with that in branches 1+2 
applying  condition (\ref{um}) and  by considering that the node is reached by formally 
taking  $n\rightarrow 0$:
\begin{eqnarray}
1+b=C\alpha_k + D\alpha_{k'} = C\beta_k + D \beta_{k'}\,.\label{kirkof1}
\end{eqnarray}

In order to write Kirchhoff rule, we use  the following  expression for the  probability current: 
\begin{eqnarray}
\bm j(\vr)={\rm Re} \left\{ \psi^\dagger(\vr) (\partial\hat H/\partial \hat \vk)\psi(\vr)
\right\}
\label{current}\end{eqnarray}
where the Hamiltonian is written in momentum space and  the operator $\hat \vk=-i\nabla$ in the continuum limit. 
In the tight-binding  problem above, $\nabla$ reduces to $\partial/\partial n$, and  the Hamiltonian  $\hat H$ is just 
the scalar dispersion $\varepsilon(\hat p)$ in branch 3, or the  Hamiltonian  matrix
 $\hat H_{1+2}$ in equation (\ref{eigen}) in branches 1+2.
If we write the Kirchhoff rule  as the following relation between the wavefunctions at the circuit node:
\begin{eqnarray}\left[
\frac{\partial\varepsilon}{\partial \hat p}\ \psi_{inc}\right]_{n\rightarrow 0^-} = 
(1,1)\cdot\left[\frac{\partial\hat H_{1+2}}{\partial \hat k}\ \psi_{t}\right]_{n\rightarrow 0^+}\,,
\label{kirkof2}
\end{eqnarray}
then, it can easily be checked that  the current $j(n)$ is conserved at the node, by virtue of 
equation (\ref{kirkof1}).  
The left multiplication by $(1,1)$ gives the sum of the currents through branches 1 and 2. 
Equation (\ref{kirkof2}) reads:  
\begin{eqnarray}
\frac{p}{m_n}(1-b) &=&
 C (1,1)\cdot\frac{\partial\hat H_{1+2}}{\partial k}\left(\begin{array}{c}\alpha_k\\ \beta_k\end{array}\right) \nonumber\\
&+& 
D (1,1)\cdot\frac{\partial\hat H_{1+2}}{\partial k'}\left(\begin{array}{c}\alpha_{k'}\\ \beta_{k'}\end{array}\right) 
\,.
\label{doisexample}
\end{eqnarray}
where the effective mass $m_n$ is defined as the ratio between the momentum, $p$, and the group velocity, $d\varepsilon(p)/dp$.
The three equations (\ref{kirkof1}) and (\ref{doisexample}) uniquely determine the amplitudes $b, C, D$.

The generalization to  two spatial dimensions is straightforward: the chain in figure \ref{tightexample} 
may be identified with the $x$ direction and is repeated identically in the perpendicular ($y$) direction. 
The unit cell label and momentum become  two-dimensional, $\bm n$ and $\bm k$, respectively. 
The interface is attained as $n_x\rightarrow 0$,  the transverse momentum component, $k_y$, is conserved.   
In equation (\ref{kirkof1}) $k$ ($k'$) is replaced by $\bm k$ ($\bm k'$) and  
the Kirchhoff rule (\ref{doisexample}) is replaced with:
\begin{eqnarray}
\frac{p_x}{m_n}(1-b) &=&
 C (1,1)\cdot\frac{\partial\hat H_{1+2}}{\partial k_x}\left(\begin{array}{c}\alpha_\vk\\ \beta_\vk\end{array}\right) \nonumber\\
&+& 
D (1,1)\cdot\frac{\partial\hat H_{1+2}}{\partial k_x'}\left(\begin{array}{c}\alpha_{\vk'}\\ \beta_{\vk'}\end{array}\right) 
\,,
\label{II}
\end{eqnarray}
ensuring the conservation of the longitudinal current $j_x$ at the node.

\section{Model for a Fe-pnictide superconductor}

A recent tight-binding model\cite{raghu2008} for the FAS band structure  assumes
two orbitals per unit  cell, $\rm d_{xz}$ and $\rm d_{yz}$.
The Hamiltonian matrix is
\begin{equation}
\hat H(\vk)=\left(\begin{array}{cc} \varepsilon_{x}- \mu\ & \varepsilon_{xy}\\
 \varepsilon_{xy} & \varepsilon_{y}- \mu\
\end{array}\right)\,,
\label{raghu}
\end{equation}
where $\mu$ denotes the chemical potential and
\begin{eqnarray}
\varepsilon_{x}&=&-2t_1\cos(k_x) -2t_2 \cos(k_y) -4t_3\cos(k_x)\cos(k_y)\nonumber \\
\varepsilon_{y}&=&-2t_2\cos(k_x) -2t_1\cos(k_y) -4t_3\cos(k_x)\cos(k_y)\nonumber \\
\varepsilon_{xy}&=&-4t_4 \sin(k_x)\sin(k_y)\,.\label{detalhes}
\end{eqnarray}
This is  analogous to branches 1 and 2 of the waveguide above, with the matrix element $\varepsilon_{xy}$ now playing the role 
of the hybridization $V(\bm k)$ between the branches 1 and 2 and $\varepsilon_{x(y)}(\bm k)$ playing the role of $\varepsilon_{1(2)}(\bm k)$.
The parameter choice $t_1=-1$, $t_2=1.3$ $t_3=t_4=-0.85$, $\mu=1.45$ reproduces the FAS band structure\cite{raghu2008}.
In the unfolded Brillouin Zone (BZ), the  Fermi surface obtained from (\ref{raghu}) has two electron pockets, centered at 
$(0,\pm\pi)$ and $(\pm\pi,0)$, and two hole pockets, centered at $(0,0)$ and $(\pi,\pi)$. 

\begin{figure}[htb]
\centerline{\includegraphics[width=9.0cm]{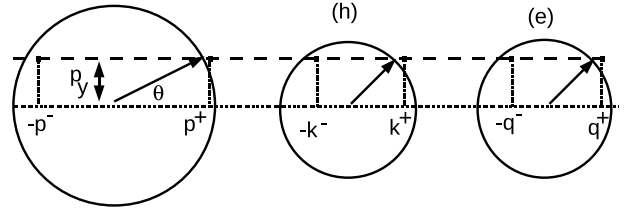}}
\caption{Schematic representation of the 
Fermi surfaces of the normal metal (left), and the superconductor's
$h$ band (middle) and $e$ band (right).
}
\label{FSs}
\end{figure}

We assume  the edge of the superconductor lying along the $y$ direction. Then, an incident electron on the interface with small $p_y$ is transmitted  
through two Fermi surface pockets: the electron pocket (``$e$ FS'')  and the  hole pocket (
``$h$ FS''). See Figure \ref{FSs}.
We here work out the Andreev reflection problem in a FS consisting of just one hole and one electron pocket. 
The generalization of the theory to the four pocket FS in the reduced BZ or to a model with more atoms per unit 
cell\cite{tesanovic,cao,kuroki} is straightforward.
We shall concentrate below on the s$^\pm$-state scenario  for superconductivity that 
has recently been suggested\cite{mazin}, and show that it produces ABS as a consequence of interference between  
transmitted waves  in the two FS pockets.
 
An  elementary excitation in the bulk superconductor with wavevector $\vk$ has the wavefunction:
\begin{equation}
\phi_\vk(\vr)=e^{i\vk\cdot\vr}\left(\begin{array}{c}u_\vk\alpha_\vk \\ u_\vk\beta_\vk \\
v_\vk\alpha_\vk \\ v_\vk\beta_\vk\end{array}\right)\,,
\label{4excite}
\end{equation}
where the coherence factors $u_\vk,v_\vk$, denoting the amplitudes of the particle and hole components, respectively, obey the Bogolubov-deGennes equations\cite{bangchoi}:
\begin{equation}
\left(\begin{array}{cc} \hat H(\vk) & \hat \Delta \\ \hat \Delta & -\hat H(\vk)
\end{array}
\right) \left(\begin{array}{c}u_\vk\alpha_\vk \\ u_\vk\beta_\vk \\
v_\vk\alpha_\vk \\ v_\vk\beta_\vk\end{array}\right)
= {\cal E}\left(\begin{array}{c}u_\vk\alpha_\vk \\ u_\vk\beta_\vk \\
v_\vk\alpha_\vk \\ v_\vk\beta_\vk\end{array}\right)\,,
\end{equation}
with $\hat \Delta  =  \Delta(\vk) {\rm diag} (1,1)$.
The superconducting gap $\Delta(\vk)$ is assumed to take on different values,
 $\Delta_h(\vk)$ and $\Delta_e(\vq)$, in the $h$  and $e$ FS, respectively.
In the s$^\pm$-state scenario,  $\Delta_e(\vk)$ and $\Delta_h(\vk)$ have  opposite signs\cite{mazin,bangchoi}.

The quasi-particle has a transverse momentum $\hbar p_y$ which is conserved. The
incident particle from the normal metal has momentum $\vp^+=\hbar(p^+,p_y)$
and the Andreev reflected hole has momentum $\vp^-=\hbar(p^-,p_y)$. 
The transmitted particle (hole) in the superconductor's $e$ band has 
momentum $\vq^+=\hbar(q^+,p_y)$ [$\vq^-=\hbar(-q^-,p_y)$]; but the 
transmitted particle (hole) in the superconductor's $h$ band has momentum
$\vk^-=\hbar(-k^-,p_y)$ [$\vk^+=\hbar(k^+,p_y)$] because the effective mass, $m_h$, of the $h$ FS
is negative and transmitted particles/holes must have positive group velocity.
See Figure \ref{FSs}.
The wavefunction for a scattering state with transverse momentum $\hbar p_y$ can be written as: 
$$
\Psi(\vr)=e^{ip_yy}\left[\psi_N(x) \theta(-x) + \psi_S(x)\theta(x)\right]
$$
where $\theta(x)$ denotes the Heaviside function.
The wave function in the normal single-band metal  has both particle ($u$) and hole ($v$) components:
\begin{equation}
\psi_N(x<0)=\left(\begin{array}c 1\\0\end{array} \right)e^{ip^+x} + 
b\left(\begin{array}{c} 1\\0\end{array} \right)e^{-ip^+x} + 
a \left(\begin{array}{c} 0\\1\end{array} \right)e^{ip^-x}\,,
\label{before}
\end{equation}
where $a$ is the Andreev reflection amplitude.
Near the Fermi level, $p^+ \approx p^- \approx p_F\sqrt{1-\left(p_y/p_F\right)^2}$,
where $\hbar p_F$ denotes the Fermi momentum in the normal metal,  
which has Fermi velocity $v_F=\hbar p_F/m_n$. 
The transmitted quasi-particle into the superconductor is a linear superposition of Bloch states
of the form (\ref{4excite})in the two bands:
\begin{eqnarray}
e^{ip_yy}\psi_S(x>0)&=& C \phi_{\vk^+}(\vr) + D \phi_{\vk^-}(\vr)\nonumber\\ &+& E\phi_{\vq^+}(\vr) + F\phi_{\vq^-}(\vr)
\label{trasm}
\end{eqnarray}

We now apply the waveguide matching conditions, at $x=0$, 
between (\ref{before}) and  (\ref{trasm}),
to the $u$ and $v$ components of the wave function, separately.
The condition for the wave function to be single valued at the node reads as:
\begin{eqnarray}
1+b= Cu_{\vk^+}\alpha_{\vk^+} + Du_{\vk^-}\alpha_{\vk^-} + Eu_{\vq^+}\alpha_{\vq^+} + Fu_{\vq^-}\alpha_{\vq^-}\,,\nonumber\\
1+b= Cu_{\vk^+}\beta_{\vk^+} + Du_{\vk^-}\beta_{\vk^-} + Eu_{\vq^+}\beta_{\vq^+} + Fu_{\vq^-}\beta_{\vq^-}\,,\nonumber\\
a= Cv_{\vk^+}\alpha_{\vk^+} + Dv_{\vk^-}\alpha_{\vk^-} + Ev_{\vq^+}\alpha_{\vq^+} + Fv_{\vq^-}\alpha_{\vq^-}\,,\nonumber\\
a= Cv_{\vk^+}\beta_{\vk^+} + Dv_{\vk^-}\beta_{\vk^-} + Ev_{\vq^+}\beta_{\vq^+} + Fv_{\vq^-}\beta_{\vq^-}\,.\nonumber\\
\label{ABCD}
\end{eqnarray}
By solving the system (\ref{ABCD}) by the determinant method, 
the amplitudes $C,D,E,F$ can be expressed as functions of $a$ and $b$, as:
\begin{eqnarray}
C&=&\frac{(1+b)\Gamma_1 +a\Gamma_2 }{\Lambda}\label{gama1}\,,\\
D&=&\frac{(1+b)\Gamma_3 +a\Gamma_4 }{\Lambda}\,,\\
E&=&\frac{(1+b)\Gamma_5 +a\Gamma_6 }{\Lambda} \,,\\
F&=&\frac{(1+b)\Gamma_7 +a\Gamma_8 }{\Lambda} \,,\label{gamas}
\end{eqnarray}
where $\Lambda$ is the determinant of the system (\ref{ABCD}) and reads:
\begin{equation}
\Lambda = \left| \begin{array}{cccc} u_+\alpha_+ & u_-\alpha_- & u_+'\alpha_+' & u_-'\alpha_-'\\
u_+\beta_+ & u_-\beta_- & u_+'\beta_+' & u_-'\beta_-'\\
v_+\alpha_+ & v_-\alpha_- & v_+'\alpha_+' & v_-'\alpha_-'\\
v_+\beta_+ & v_-\beta_- & v_+'\beta_+' & v_-'\beta_-'\\
\end{array} \right|\,,
\label{lambda}
\end{equation}
and the coefficients $\Gamma_i$ are obtained from Cramer's rule.

We now define:
\begin{eqnarray}
\Theta(\vk) &=& \frac{m_n}{p_+}(1,1)\cdot \left(u\frac{\partial \hat H}{\partial k_x} + v \frac{\partial \hat \Delta}{\partial k_x}\right) 
\left(\begin{array}{c} \alpha \\ \beta\end{array} \right)\,,\\
\Phi(\vk) &=& \frac{m_n}{p_-}(1,1)\cdot \left(v\frac{\partial \hat H}{\partial k_x} - u \frac{\partial \hat \Delta}{\partial k_x}\right) 
\left(\begin{array}{c} \alpha \\ \beta\end{array} \right)\,,
\end{eqnarray}
and write condition (\ref{II}) for this case as:
\begin{eqnarray} 
1-b &=& C\Theta(\vk_+) + D \Theta(\vk_-) + E \Theta(\vq_+) + F\Theta(\vq_-)\,,\nonumber \\
a &=& C\Phi(\vk_+) + D \Phi(\vk_-) + E \Phi(\vq_+) + F\Phi(\vq_-)\,. \nonumber  \\
\label{correntes}
\end{eqnarray}

In order to simulate interface disorder, a potential barrier $U\delta(x-\epsilon)$ is assumed
in the normal metal  ($\epsilon<0$)  and the limit  $\epsilon\rightarrow 0^-$
is taken\cite{eu}.
We now show that  the effect of the barrier amounts to making the replacement:  
\begin{eqnarray}
1-b &\rightarrow & 1-b-2iZ(1+b)p_F/p^+ \label{breplace}\\
a &\rightarrow & a(1-2iZp_F/p^+)\label{areplace}
\end{eqnarray}
on the right-hand side of equation (\ref{correntes}), 
where the dimensionless barrier parameter\cite{blonder}
$Z=U/\hbar v_F$.
To see this, we write the  wave function in the normal single-band metal with 
both particle and hole components:
\begin{equation}
\psi_N(x\leq\epsilon)=\left(\begin{array}c 1\\0\end{array} \right)e^{ip^+x} + 
b\left(\begin{array}{c} 1\\0\end{array} \right)e^{-ip^+x} + 
a \left(\begin{array}{c} 0\\1\end{array} \right)e^{ip^-x}\,,
\end{equation}
and
\begin{eqnarray}
\psi_N(\epsilon<x<0)&=&\tilde\alpha\left(\begin{array}c 1\\0\end{array} \right)e^{ip^+x} + 
\tilde\beta\left(\begin{array}{c} 1\\0\end{array} \right)e^{-ip^+x}\nonumber\\ 
&+& \tilde\gamma\left(\begin{array}c 0\\1\end{array} \right)e^{ip^-x} + 
\tilde\delta\left(\begin{array}{c} 0\\1\end{array} \right)e^{-ip^-x}\,.
\nonumber \\
& & \label{justbefore}
\end{eqnarray}
The matching conditions at $x=\epsilon<0$ give the equations:
\begin{equation}
\psi_N(\epsilon^-)=\psi_N(\epsilon^+)\,,
\end{equation}
\begin{equation}
-\frac{\hbar^2}{2m_n}\left[ \psi_N'(\epsilon^+)- \psi_N'(\epsilon^-)\right] 
+U \psi_N(\epsilon) =0\,.
\end{equation}
Taking the limit $\epsilon\rightarrow 0^-$ we obtain:
\begin{eqnarray}
\tilde\alpha+\tilde\beta &=& 1+b\,,\label{a+b}\\
\tilde\alpha-\tilde\beta &=& \frac{2m_nU}{i\hbar^2p^+} (1+b) +1-b\,,\label{a-b}\\
\tilde\gamma +\tilde\delta &=& a\,,\label{g+d}\\
\tilde\gamma -\tilde\delta &=& \left( \frac{2m_nU}{i\hbar^2p^-} + 1 \right) a\label{g-d}
\,.
\end{eqnarray}
The waveguide matching conditions must be applied between (\ref{justbefore}) and 
(\ref{trasm}) at $x=0$.
But equations (\ref{a+b}) through (\ref{g-d}) allow the elimination of the amplitudes
$\tilde\alpha, \tilde\beta, \tilde\gamma, \tilde\delta$ altogether, finally showing
that the replacements (\ref{breplace})-(\ref{areplace}) have to be done in equation (\ref{correntes}).

The values of the Andreev and normal reflection amplitudes, $a$ and $b$,
can be obtained by solving the linear system (\ref{ABCD}) and (\ref{correntes}). 
 Introducing
\begin{eqnarray}
\zeta_{11} &=& \Gamma_1 \Theta(\vk_+) + \Gamma_3 \Theta(\vk_-) + \Gamma_5 \Theta(\vq_+) + \Gamma_7 \Theta(\vq_-)\nonumber\\
\zeta_{12} &=& \Gamma_2 \Theta(\vk_+) + \Gamma_4 \Theta(\vk_-) + \Gamma_6 \Theta(\vq_+) + \Gamma_8 \Theta(\vq_-)\nonumber\\
\zeta_{21} &=& \Gamma_1 \Phi(\vk_+) + \Gamma_3 \Phi(\vk_-) + \Gamma_5 \Phi(\vq_+) + \Gamma_7 \Phi(\vq_-)\nonumber\\
\zeta_{22} &=& \Gamma_2 \Phi(\vk_+) + \Gamma_4 \Phi(\vk_-) + \Gamma_6 \Phi(\vq_+) + \Gamma_8 \Phi(\vq_-)\,,\nonumber\\
\end{eqnarray}
we obtain:
\begin{eqnarray}
a=\frac{2\zeta_{21}/\Lambda}
{\left(1+2iZ+\frac{\zeta_{11}}{\Lambda}\right)\left(1-2iZ-\frac{\zeta_{22}}{\Lambda}\right)+ 
\frac{\zeta_{12}\zeta_{21}}{\Lambda^2}}\,,\label{a}
\end{eqnarray}
and
\begin{eqnarray}
b=\frac{\left(1-2iZ-\frac{\zeta_{11}}{\Lambda}\right)\left(1-2iZ-\frac{\zeta_{22}}{\Lambda}\right)-\frac{\zeta_{12}\zeta_{21}}{\Lambda^2}
}
{\left(1+2iZ+\frac{\zeta_{11}}{\Lambda}\right)\left(1-2iZ-\frac{\zeta_{22}}{\Lambda}\right)+ \frac{\zeta_{12}\zeta_{21}}{\Lambda^2}
}\,.\label{b}
\end{eqnarray}
The contribution of this scattering state to the differential  conductance is given by:
\begin{eqnarray}
g_s=1+|a|^2-|b|^2\,.
\label{gs}
\end{eqnarray}

The normal state conductance, $g_n=1-|b_n|^2$, is obtained when $\Delta(\vk)=0$. 
Experimentally,  the  integral of $g_s$ (or $g_n$)  over
the transverse momenta of the incident electrons, 
$\sigma_S = \int g_s dp_y$ (or $\sigma_N = \int g_n dp_y$ ) is measured\cite{tanaka}.
We define the integrated relative  differential conductance as $\sigma_S / \sigma_N$.

\section{Results}

The conductances $\sigma_S$ and $g_s({\cal E})$ have been calculated 
from the above theory using  the model (\ref{raghu})-(\ref{detalhes}). 
We discuss the results below.
In the calculations, the normal metal is assumed to have  Fermi wavevector 
$p_F=\pi$ and velocity $v_F=1.83$.

\begin{figure}[htb]
\vspace{0.3cm}
\centerline{\includegraphics[width=7.0cm]{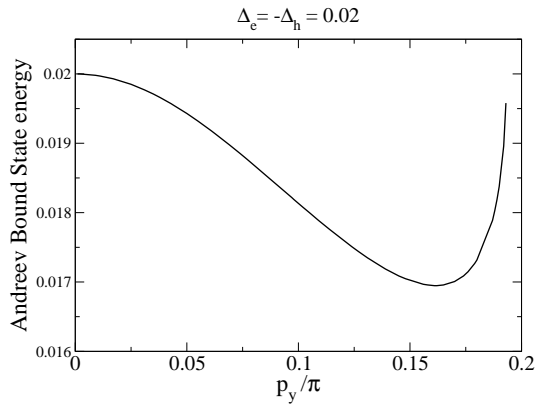}}
\caption{Energy of the surface Andreev bound state as function of the transverse momentum.
$\Delta_e=-\Delta_h=0.02$.}
\label{ABSfigure}
\end{figure}

When $\Lambda=0$ the reflection amplitudes, 
hence $g_s$, become independent of the barrier parameter $Z$ and this is
precisely the condition for the occurrence of Andreev bound
states\cite{tanaka,golubov,note}. 
Figure \ref{ABSfigure} shows
the energy of the Andreev bound state as function of the transverse momentum. Contrary to the usual case
of single band non-conventional superconductors,  the ABS energy is nonzero.
It has a non monotonic dependence on $p_y$ and, for $p_y=0$, it coincides with
 ${\rm min}\left(|\Delta_h|,|\Delta_e|\right)$.
The dispersion of the ABS energy is in qualitative agreement with the  results of Ref \cite{ghaemi}.

\begin{figure}[htb]
\vspace{0.5cm}
\centerline{\includegraphics[width=7.0cm]{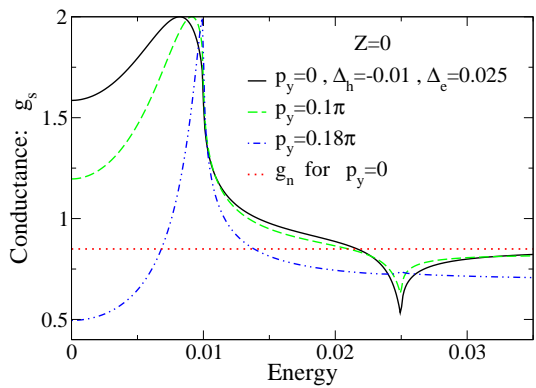}}
\caption{Conductance $g_s$ as function of incident electron energy for three different
values of the transverse momentum, for a clean $(Z=0)$ interface. The superconductor
bands are modeled by equations (\ref{raghu})-(\ref{detalhes}). The normal conductance is shown for comparison.}
\label{clean}
\end{figure}

\begin{figure}[htb]
\vspace{0.5cm}
\centerline{\includegraphics[width=7.0cm]{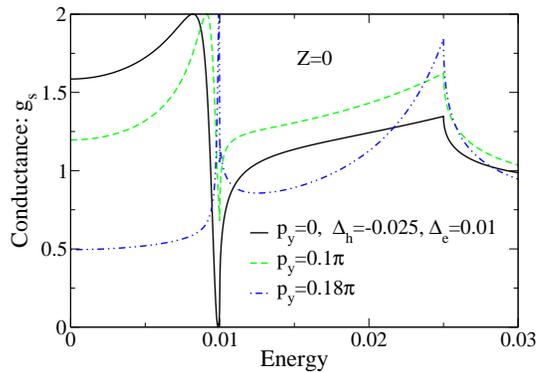}}
\vspace{0.3cm}
\caption{Conductance $g_s$ as function of incident electron energy for three different
values of the transverse momentum, for a clean $(Z=0)$ interface.
$\Delta_e=0.01, \Delta_h=-0.025$.}
\label{hmaiore}
\end{figure}

Figure \ref{clean} shows the conductance $g_s$ as function of incident electron energy above the Fermi level,
for a clean ($Z=0$) interface in the case where $|\Delta_h|<|\Delta_e|$. When the transverse momentum increases,
$g_s$ becomes more strongly peaked near the energy of the ABS.
In the case where $|\Delta_h|>|\Delta_e|$ there is a destructive interference effect leading to a zero, at normal incidence,
in the conductance,  as shown in Figure \ref{hmaiore}.

\begin{figure}[htb]
\vspace{0.6cm}
\centerline{\includegraphics[width=7.0cm]{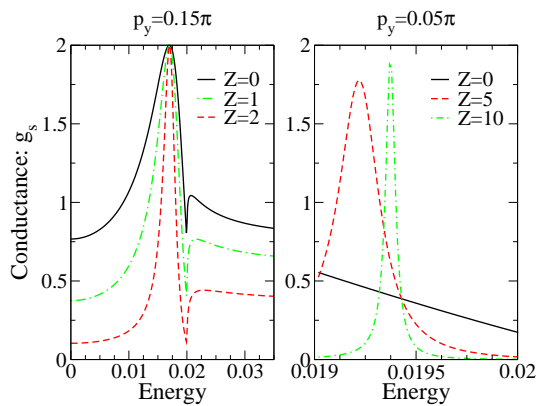}}
\caption{The conductance $g_s$ is independent of the barrier strength, $Z$, 
at the energy of the ABS. The latter can be checked from Fig. \ref{ABSfigure}.
This is shown for two transverse momenta: $p_y=0.15\pi$ (left); $p_y=0.05\pi$ (right).
$\Delta_e=-\Delta_h=0.02$.}
\label{bound}
\end{figure}

\begin{figure}[htb]
\vspace{0.7cm}
\centerline{\includegraphics[width=8.0cm]{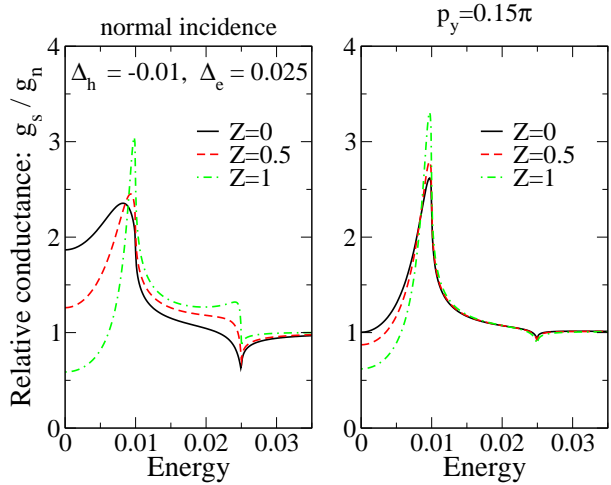}}
\caption{Relative conductance $g_s/g_n$ for a disordered interface, at two different values of transverse momentum.
$\Delta_h=-0.01, \Delta_e=0.025$.
}
\label{hmenoreZ}
\end{figure}

\begin{figure}[htb]
\vspace{0.7cm}
\centerline{\includegraphics[width=7.0cm]{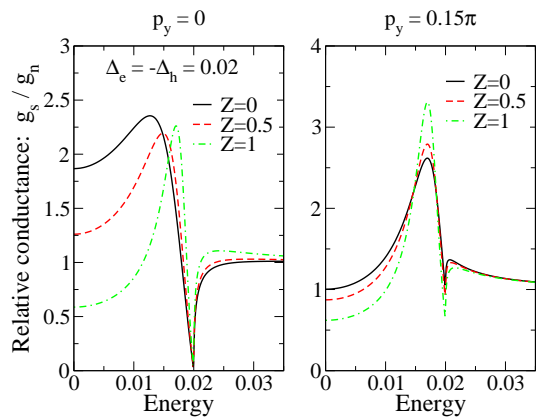}}
\vspace{0.3cm}
\caption{
Behavior of the  relative conductance for different disorder values, $Z$.
Destructive interference effects cause a zero in the conductance.
$\Delta_e=-\Delta_h=0.02$.}
\label{heiguais}
\end{figure}

The effect of interface disorder is shown in Figures \ref{bound}, \ref{hmenoreZ} and 
\ref{heiguais}. 
Under increasing disorder, conductance peaks appear closer to the energy of ABS. The conductance $g_s$ is independent
of the disorder parameter, $Z$, at the ABS energy, therefore, all the conductance curves $g_s(E)$, for different $Z$ values, intercept
at the same point, as shown in Figure \ref{bound}. As disorder increases the conductance
tends to decrease and, therefore, the normal state conductance decreases as $Z$ increases.
Since at the ABS
the conductance $g_s$ is independent
of the disorder parameter, $Z$, peaks appear in the relative conductance at the ABS,
which become more pronounced as $Z$ gets larger.
The relative conductance $g_s/g_n$ is plotted for two ratios of the gap parameters in
Figures \ref{hmenoreZ} and \ref{heiguais}, illustrating the positions of the peaks.
The case when $\Delta_h=- \Delta_e$ is shown in Figure \ref{heiguais}. It is also seen, once again,
that destructive interference effects
in the superconductor between the $e$ and $h$ bands cause the conductance to vanish 
at normal incidence.

\begin{figure}[htb]
\vspace{0.9cm}
\centerline{\includegraphics[width=7.0cm]{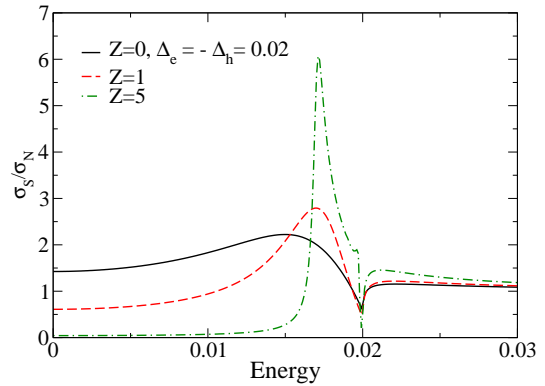}}
\vspace{0.3cm}
\caption{
Behavior of the  integrated relative differential conductance, $\sigma_S/\sigma_N$, for different disorder values, $Z$.
$\Delta_e=-\Delta_h=0.02$.}
\label{iguaistotal}
\end{figure}

Since the system is two-dimensional we have to integrate over the possible
incident angles, or over $p_y$.
The integrated relative conductance is shown in Figure \ref{iguaistotal} for 
three interface disorder strengths.
The peak structure  for large values of the barrier strength
reveals the existence of ABS's. 

\begin{figure}[htb]
\vspace{0.3cm}
\centerline{\includegraphics[width=7.0cm]{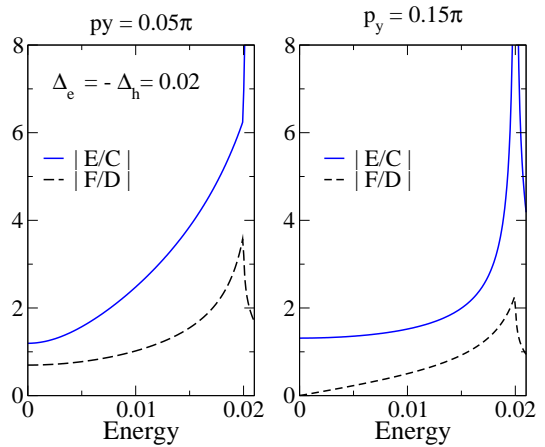}}
\caption{
The ratio of the moduli of amplitudes  $|E/C|$ and $|F/D|$ as a function of energy.
The transverse momentum $p_y=0.05\pi$ (left panel) and $p_y=0.15\pi$ (right panel). $\Delta_e=-\Delta_h=0.02$.
}
\label{comparacao}
\end{figure}

In a recent preprint\cite{golubov}, a theory is provided that  qualitatively predicts
the interference effects in the multiband superconductor, namely the suppression of conductance and
the appearance of ABS, in agreement with our findings.
In Ref \cite{golubov}, the wave function in the superconductor is written in
the same form as  equation (\ref{trasm}) but 
it is assumed that the ratios of the amplitudes  $E/C$ and $F/D$ are equal
(to a phenomenological parameter $\alpha$ introduced in Ref \cite{golubov}).
Using our waveguide theory for the interface matching conditions, we find that the ratios
$E/C$ and $F/D$ are different, as can be seen from figure \ref{comparacao}.

\section{Summary}

We have introduced a generalization of the
quantum waveguide theory to determine the appropriate boundary conditions for
the wave function at the interface between a  normal metal and a multiband superconductor.
We have shown that resonant transmission and destructive interference
effects occur  in the sign-reversed scenario for pnictide superconductors.
 Unlike  other unconventional superconductors, 
Andreev bound states at finite energies are brought about by these interference effects. 

On the experimental side, polycrystalline samples have been used so far. The results obtained above describe an interface parallel to the nearest Fe-Fe bonding. Therefore, experiments with single crystals are highly desirable. 
If the edge of the sample is such that the conservation of the transverse momentum $p_y$ intercetps only one FS pocket,  existing one band theories apply. 
The above quantum waveguide theory can in principle be used  
to describe other MBS, such as the heavy-fermion materials\cite{eu}.

\section*{Acknowledgments}

We would like to thank V\'{\i}tor R. Vieira for a discussion and comments on the manuscript.



\begin{thebibliography}{99}
\bibitem{note} The limit $\Lambda\rightarrow 0$ leads to zeros in the parameters $\zeta_{i j}$. 
L'H\^opital's rule must be observed while taking limit $\Lambda\rightarrow 0$
in equations  (\ref{a})-(\ref{b}).
\end{thebibliography}
\end{document}